\documentclass[aip,cha,reprint]{revtex4-1}

\usepackage[pdftex]{graphicx}
\usepackage{amssymb,amsfonts,amsmath}

\begin{document}


\title{Coexistence of synchrony and incoherence in oscillatory media \\
  under nonlinear global coupling} 



\author{Lennart Schmidt}
\affiliation{Physik-Department, Nonequilibrium Chemical Physics, Technische Universit\"{a}t M\"{u}nchen,
  James-Franck-Str. 1, D-85748 Garching, Germany}
\affiliation{Institute for Advanced Study - Technische Universit\"{a}t M\"{u}nchen,
  Lichtenbergstr. 2a, D-85748 Garching, Germany}

\author{Konrad Sch\"{o}nleber}
\affiliation{Physik-Department, Nonequilibrium Chemical Physics, Technische Universit\"{a}t M\"{u}nchen,
  James-Franck-Str. 1, D-85748 Garching, Germany}

\author{Katharina Krischer}
\email[]{krischer@tum.de}
\affiliation{Physik-Department, Nonequilibrium Chemical Physics, Technische Universit\"{a}t M\"{u}nchen,
  James-Franck-Str. 1, D-85748 Garching, Germany}

\author{Vladimir Garc\'{i}a-Morales}
\affiliation{Physik-Department, Nonequilibrium Chemical Physics, Technische Universit\"{a}t M\"{u}nchen,
  James-Franck-Str. 1, D-85748 Garching, Germany}
\affiliation{Institute for Advanced Study - Technische Universit\"{a}t M\"{u}nchen,
  Lichtenbergstr. 2a, D-85748 Garching, Germany}

\date{\today}

\begin{abstract}
We report a novel mechanism for the formation of chimera states, a peculiar
spatiotemporal pattern with coexisting synchronized and incoherent
domains found in ensembles of identical oscillators.
Considering Stuart-Landau oscillators we demonstrate
that a nonlinear global coupling can induce this symmetry breaking. We find chimera
states also in a spatially extended system, a modified
complex Ginzburg-Landau equation. This theoretical prediction is
validated with an
oscillatory electrochemical system, the electrooxidation of silicon,
where the spontaneous formation of chimeras is observed without
any external feedback control.
\end{abstract}


\maketitle 

\begin{quotation}
In the 17th century Christiaan Huygens was the first who encountered
the phenomenon of synchronization, when watching two coupled pendulum
clocks adjusting their oscillation phase to each other. Since then, a
variety of systems exhibiting synchronization were studied, e.g. the
flashing of fireflies or networks of pacemaker cells keeping our heart
beating in time. In these systems the key aspect is that nonidentical
oscillating elements, as nature is never perfect, with a distribution
of natural frequencies become synchronized due to the mutual coupling.
In contrast, in 2002 \textit{Kuramoto \& Battogtokh} \cite{Kuramoto_NPCS_2002} found the
opposite phenomenon: a perfect symmetric system of identical
oscillators coupled via a nonlocal coupling (i.e. a coupling that
somehow decreases with the distance between two oscillators) may
undergo a transition to a state, where a synchronized group of
oscillators coexists with an unsynchronized one. This situation was
later named chimera state, since the chimera was, according to Greek
mythology, composed of the parts of different animals. The nonlocality
of the coupling is believed to be indispensable for the formation of
chimera states. However, in the present article we show that this is a
misbelief, as we found chimera states under solely global
coupling. Global coupling means that each individual oscillator
couples to the mean field of all oscillators. In our case the mean field is a
nonlinear function of the state variables of each oscillator.
\end{quotation}


\section{Introduction}

An oscillatory medium experiencing a global
coupling or feedback mechanism may evolve towards a domain-like structure
called a cluster state, in which each domain oscillates uniformly with
a defined phase difference to the others \cite{Mikhailov_PhysicsReports_2006}.  Several theoretical
studies on nonlocally coupled oscillatory systems predicted a strange
domain-type pattern, called a chimera state, where some
domains are perfectly synchronized, but others oscillate spatially
incoherently \cite{Kuramoto_NPCS_2002, AbramsStrogatz_PRL_2004, Shima_PRE_2004, Martens_PRL_2010,
    Sethia_PRL_2008, Abrams_PRL_2008, Schoell_PRL_2011,
    Omelchenko_PRE_2012, Nkomo_PRL_2013, Omelchenko_PRL_2013}. Chimera states might be of importance for some peculiar
observations in different disciplines, such as the unihemispheric
sleep of animals \cite{Rattenborg_NBRev_2000, Mathews_Ethology_2006}, the need for synchronized bumps in
otherwise chaotic neuronal networks for signal propagation \cite{Vogels_ARN_2005} and
the existence of turbulent-laminar patterns in a Couette flow \cite{Barkley_PRL_2005}.
Very recently, the existence of chimera states could be validated in
two pioneering experiments with chemical \cite{Tinsley_Nature_2012} and optical oscillators
\cite{Hagerstrom_Nature_2012}. Both
experiments involved a specifically designed feedback algorithm to
generate the specific nonlocal coupling. Subsequently, 
chimera states could be realized in systems of mechanical
\cite{Martens_PNAS_2013} and electrochemical \cite{Wickramasinghe_PONE_2013} oscillators.
However, experimental evidence of the
spontaneous formation of chimera states without the control from
outside is still missing.

In this Article we demonstrate, both theoretically and experimentally,
that also under a strictly global coupling, if being nonlinear,
a coexistence of synchrony and asynchrony can be found. We start with an ensemble of
Stuart-Landau oscillators interacting solely via a nonlinear global
coupling. An initially random distribution splits
for given parameters into two groups, one being synchronized and the
other one being desynchronized.
We then discuss spatially extended oscillatory media.
We show that a modified complex
Ginzburg-Landau equation with nonlinear global coupling, originally proposed to
explain special cluster patterns observed during the oscillatory
electrooxidation of silicon in fluoride containing electrolytes \cite{Miethe_PRL_2009,
  GarciaMorales_PRE_2010}, describes a transition from cluster patterns to
a state with coexisting synchronized and incoherent domains. The results
are indeed confirmed experimentally with the oscillating Si-system, where the
separation of the electrode into coherently and incoherently
oscillating domains occurs spontaneously and without external
feedback control. Most remarkably, the incoherent region does not
contain any amplitude defects. All these are essential properties of
chimera states and we conclude that we have found a novel mechanism
to this symmetry-breaking state. Moreover, since a global coupling is frequently
encountered, chimera states might exist in many more systems than
anticipated so far.

\section{Results and discussion}
\subsection{Chimera states in an ensemble of Stuart-Landau oscillators}

First, we consider an ensemble of $N$ Stuart-Landau oscillators
\cite{Kuramoto_2003} under nonlinear global coupling

\begin{align}
  \frac{\mathrm d}{\mathrm dt} W_j &= W_j - (1 + i c_2) \left| W_j
  \right|^2 W_j  \notag \\
  &\quad - (1 + i \nu) \left< W \right> + (1 + i c_2) \left<
    \left| W \right|^2 W \right> \ ,
\label{eq:SL_ensemble}
\end{align}

where $j \in \left[ 1,N \right]$ labels each individual oscillator and
$\left< W \right> = \sum_{k=1}^N W_k / N$ and $\left< \left| W \right|^2 W
\right> = \sum_{k=1}^N \left| W_k \right|^2 W_k/N$ denote ensemble averages.
The first term on the right hand side is the linear instability
leading to oscillations, while their magnitudes are controlled by the cubic term. The last
two terms represent the nonlinear global coupling. 
Taking the ensemble average on both sides of Eq.~\eqref{eq:SL_ensemble} yields $\mathrm d\left< W
\right> / \mathrm dt = - i \nu \left< W \right>$ and thus $\left<
  W \right> = \eta \exp(-i \nu t)$, i.e. the average $\left< W
\right>$ exhibits conserved harmonic oscillations with amplitude
$\eta$ and frequency $\nu$. As we will see later, this
conservation law strongly influences the dynamics. Altogether we have
three parameters, namely $c_2$, $\nu$ and $\eta$. 

\begin{figure}[ht]
  \centering
  \includegraphics[width=75mm]{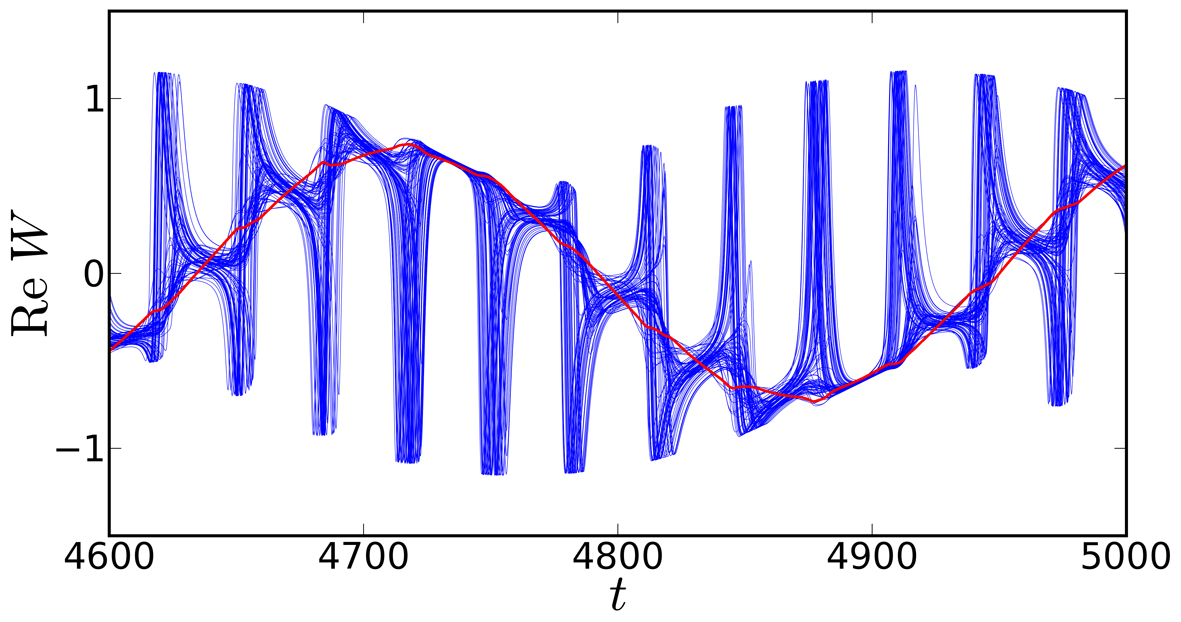}
  \caption{Time series for the real parts
  of $W_j$ for all oscillators are shown. The synchronized oscillators
(red, light gray) perform collective and nearly harmonic oscillations, while the
asynchronous oscillators (blue, dark gray) exhibit incoherent and irregular
dynamics. For parameters see text.}
  \label{fig:SL_chimera}
\end{figure}

We numerically solved Eq.~\eqref{eq:SL_ensemble} (for details see
the Appendix).
Starting from a random distribution, for $c_2 = -0.6$, $\nu = 0.02$,
$\eta = 0.7$ and $N = 1000$ the ensemble splits into two groups as depicted in
Fig.~\ref{fig:SL_chimera}, where we show the real parts of $W_j$ for
all oscillators. One group is synchronized (red, light gray) and the other group is
desynchronized (blue, dark gray). The synchronized oscillators perform collective
and nearly harmonic oscillations, while the asynchronous ones exhibit
incoherent and irregular motions. Although the oscillations of this
latter group present a seemingly regular spiking, the oscillators in
this group are strongly uncorrelated both in time and in their
simultaneous amplitudes and phases. We observe strong amplitude
fluctuations in the incoherent group, as this is also the case for
chimeras found in a nonlocally coupled system in a parameter region,
where the weak-coupling approximation does not apply \cite{Sethia_PRE_2013}. No interchange of oscillators
between the two groups occurs.

The regularity of the spiking can be explained by a second time-scale
inherent in the system.
As discussed for the continuous system in \cite{GarciaMorales_PRE_2010}, in the parameter regime, where
clustering occurs, the nonlinear global coupling leads to two dominant
time-scales: the frequency $\nu$ of the oscillation of the spatial average
and a frequency, which may be called the cluster frequency. The
contribution to the oscillations in the two phases at this cluster
frequency show a phase shift of $\pi$ (between the two phases). The
time-scale of the regular spiking in Fig.~\ref{fig:SL_chimera}
is given by the cluster frequency described above as the clustering
mechanism leads to the separation into the two groups. Interestingly,
chimera states found in an electrochemical experiment with individual
electrodes arranged on a ring and coupled nonlocally exhibit a
similar spiking behaviour: the desynchronized oscillating
elements drift some time with the mean-field, interrupted by fast
$2\pi$ phase slips \cite{Wickramasinghe_PONE_2013}.

In essence we have found the coexistence of synchrony and asynchrony,
i.e. a kind of a chimera state, evolving under a solely
global coupling. This contradicts the assumption that a nonlocal
coupling is indispensable for the occurrence of these
states. Contrarily to the findings in
\cite{Wolfrum_PRE_2011}, the chimera state is stable independently of
the population size and forms spontaneously \cite{Omelchenko_PRL_2008}
from a random distribution. The co-existing synchronized state is
unstable, which is also the case for the chimera states described in
\cite{Omelchenko_PRL_2008} and in \cite{Sethia_PRE_2013}. In the
latter work it is argued, that this is connected with
strong fluctuations of the amplitude in the incoherent region, as they did not consider
the weak-coupling limit. The type of chimera states found here are absent under
linear global coupling \cite{Nakagawa_PhysicaD_1994}. 
Note, however, that linear global coupling may induce other types
of chimera states, also in an ensemble of Stuart-Landau oscillators involving large
amplitude variations \cite{Daido_PRL_2006}, or in a globally coupled map lattice
\cite{Kaneko_PhysicaD_1990}. The former state is
also found in our model, Eq.~\eqref{eq:SL_ensemble}, and will be discussed elsewhere.


\subsection{Transition to a chimera state in the modified CGLE}

In order to describe experiments on an spatially extended oscillatory
medium, we consider now a modified complex Ginzburg-Landau
equation (MCGLE) \cite{Miethe_PRL_2009, GarciaMorales_PRE_2010}, 

\begin{align}
  \partial_t W &= W + (1 + i c_1) \nabla^2 W - (1 + i c_2)
  \left| W \right|^2 W \notag \\
  &\quad - (1 + i \nu) \left< W \right> + (1 + i c_2) \left<
    \left| W \right|^2 W \right> \ .
\label{eq:MCGLE}
\end{align}

Here $W(\mathbf r,t)$ is the complex order parameter describing the dynamical state at each
point $\mathbf r = (x,y)$ at time $t$ and $\left< \dots \right>$ now denotes the
spatial average. 
The original complex Ginzburg-Landau equation without the nonlinear global coupling is a
generic model for systems at the onset of oscillations and has a wide
range of applications
\cite{AransonKramer_RevModPhys_2002}. 
The MCGLE, Eq.~\eqref{eq:MCGLE}, was proposed to
explain experimental results of the electrooxidation of $n$-Si(111)
under illumination \cite{Miethe_PRL_2009}. In fact, the emergence of subharmonic cluster patterns
in the oxide-layer thickness at the silicon-electrolyte interface can
successfully be described by Eq.~\eqref{eq:MCGLE} \cite{GarciaMorales_PRE_2010}. An important
experimentally observed feature is a nearly harmonic oscillation of
the spatially averaged oxide-layer thickness. This is captured by the
conservation law for the homogeneous mode $\left< W \right> = \eta
\exp(-i \nu t)$ in the theory \cite{GarciaMorales_PRE_2010}.

\begin{figure}[t]
  \centering
  \includegraphics[width=85mm]{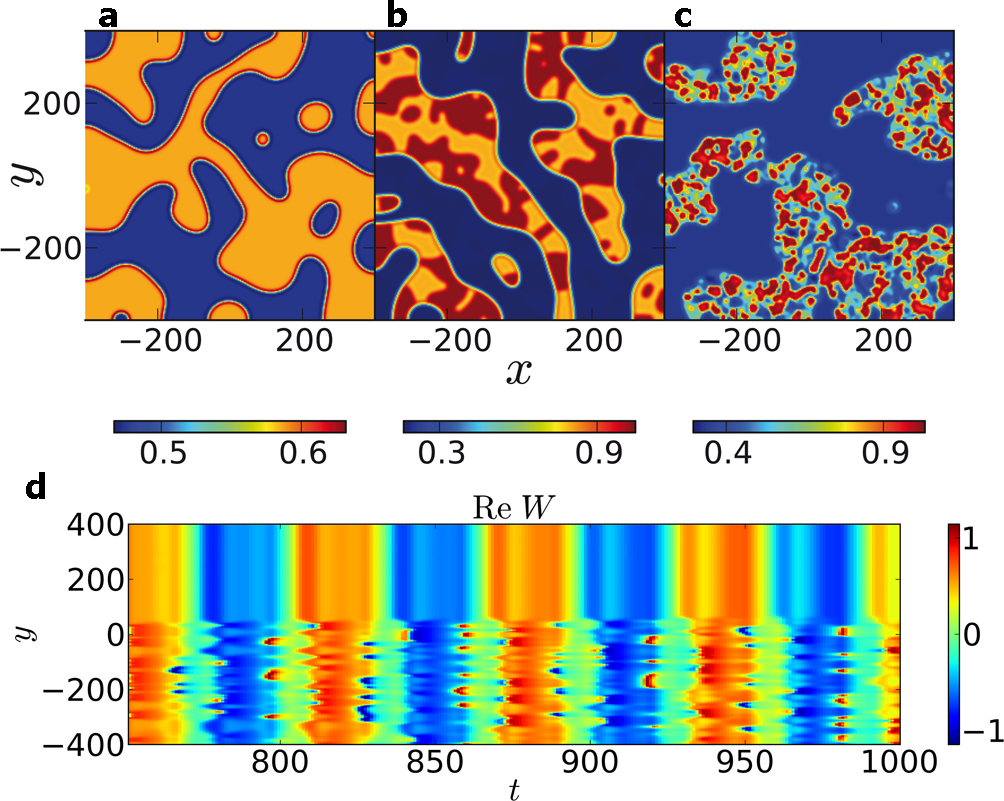}
  \caption{(\textbf{a}) - (\textbf{c}) Snapshots of the three cluster-states.
    Shown is the real part of the complex order parameter $\mathrm{Re}
  \ W$, calculated from Eq.~\eqref{eq:MCGLE}, indicating the dynamical states of each local
  oscillator. (\textbf{a}) Two-phase clusters obtained for parameter
  $c_2 = -0.7$. Both phases are homogeneous. (\textbf{b})
  Subclustering at $c_2 = -0.67$. In this case one phase is 
  homogeneous, while the other one splits again into two-phase
  clusters. (\textbf{c}) Two-dimensional chimera state found for $c_2 =
  -0.58$. The inhomogeneous phase shows strongly incoherent
  dynamics. (\textbf{d}) Temporal evolution of the real part of $W$ in a cut along the
  $y$-direction at $x=0$ in (\textbf{c}). Perfectly synchronized motion coexists with
  asynchronous behaviour, separated by a sharp boundary.}
\label{fig:W_real}
\end{figure}

We numerically solved Eq.~\eqref{eq:MCGLE} (for details see the Appendix) for fixed parameters $c_1 = 0.2$, $\nu = 0.1$
and $\eta = 0.66$. For appropriate values of $c_2$ the system
splits into two phases, as presented in Fig.~\ref{fig:W_real}a for $c_2 =
-0.7$.

The specific interaction between these two phases via the nonlinear
global coupling leads to a symmetry-breaking transition, as we will
show in the following. 
Let us call the two phases A and B, respectively. 
Simulations show that the system evolves according to a minimization
of the interface between A and B. This leads to a demixing of the
phases. As the diffusional coupling between A and B acts only near the
boundaries, for large domain sizes it can be neglected. Under this
assumption, the dynamics in each phase is governed by

\begin{align}
  \partial_t W_X(\mathbf r, t) &= W_X(\mathbf r,t) + (1 + i c_1)
  \nabla^2 W_X(\mathbf r,t) \notag \\
 - (1 + i c_2) &\left| W_X(\mathbf r,t) \right|^2 W_X(\mathbf r,t)
  + Z(W_A, W_B) \ ,
\end{align}

where X = A, B and $Z(W_A, W_B)$ is the coupling between A and B and
has to be determined. Exploiting the conservation law for the
homogeneous mode one finds

\begin{align}
  Z(W_A, W_B) &= - (1 + i \nu) \eta \exp(- i \nu t)
  \notag \\ 
   + (1 + i c_2) &\frac 12 \left( \left< \left| W_A \right|^2 W_A \right>
    + \left< \left| W_B \right|^2 W_B \right> \right) \ .
\end{align}

We can further write for the spatial averages over phases A and B $R_A \exp \left( -i\alpha \right) \equiv \left< \left| W_A
  \right|^2 W_A \right>$ and $R_B \exp \left( -i\beta \right) \equiv \left< \left| W_B
  \right|^2 W_B \right>$, respectively, and
$K \exp \left( i\gamma \right) \equiv (1+ic_2)/2$, where $\gamma =
\gamma(c_2)$. With the phase difference $\Delta \phi \equiv \beta -
\alpha$ between A and B, one can now
show that the intra-group coupling differs from the inter-group coupling. Note that $\Delta \phi$ is
generally unequal to $\pi$ as we are dealing with subharmonic
two-phase clusters \cite{Miethe_PRL_2009}. One obtains in terms
of $\alpha$

\begin{align}
  Z(W_A, W_B) &= - (1 + i \nu) \eta e^{- i \nu t}
  \notag \\ 
  & + KR_A e^{i (\gamma - \alpha)} + KR_B
  e^{i (\gamma - \Delta \phi - \alpha)} \ .
\end{align}

We see that phases A and B experience each a different influence from the
intra- and inter-group couplings. This is not due to a difference in
coupling strength defined a priori, but is the result of the
intrinsic dynamics causing the phase difference. 
As studies of two subpopulations in refs. \cite{Abrams_PRL_2008,
  Tinsley_Nature_2012} with global intra- and inter-group couplings of 
different strength show the existence of chimera
states, we conclude that the similar situation arising here renders the
emergence of chimeras possible.
The coupling can be tuned with the parameter $c_2$, where the
influence is different on inter- and intra-group coupling if
$\Delta \phi$ depends also on $c_2$, which is a reasonable assumption.

As presented in Fig.~\ref{fig:W_real}, we
find three remarkable, stable states. As already mentioned, for $c_2 = -0.7$, Fig.~\ref{fig:W_real}a,
we observe two-phase clusters. By changing to $c_2 = -0.67$, shown in
Fig.~\ref{fig:W_real}b, one finds 
A being homogeneous and B exhibiting two-phase clusters as a
substructure. Finally we observe a chimera state for $c_2
= -0.58$, where B becomes turbulent. This is depicted in
Fig.~\ref{fig:W_real}c. All these states were also found in
ref.~\cite{Tinsley_Nature_2012}, but there the two subpopulations were
man-made and the system had to be initialized in a special manner.
In contrast, in our case, the system splits spontaneously into the two
groups.

To further illustrate the characteristics of the chimera state, we show
the spatio-temporal dynamics in a cut along the $y$-direction versus
time in Fig.~\ref{fig:W_real}d. It demonstrates
the separation into two parts, one being perfectly synchronized, while
the other one exhibits asynchronous behaviour. The individual
oscillators in the homogeneous region oscillate periodically, while in
the inhomogeneous region the dynamics is irregular, but still slaved
to the oscillation of the mean value $\left< W \right>$ due to the
conservation law. As in the ensemble of Stuart-Landau oscillators, the
chimera state is stable in the MCGLE.



Now we turn towards the experimental situation, which had led to the
formulation of the modified CGLE, Eq.~\eqref{eq:MCGLE}.


\subsection{Experimental validation of theoretical prediction}

The system investigated is the photoelectrochemical
dissolution of n-type doped silicon in fluoride containing
electrolytes. Here the silicon sample is oxidized electrochemically
via the following dominant reaction \cite{Memming_SurfScience_1966}: 

\begin{align}
  \mathrm{Si + 4 H_2O + \nu_{VB} h^+} &\mathrm{\rightarrow Si\left(OH\right)_4 +
    4H^+ + (4 - \nu_{VB}) e^-} \notag \\
  \mathrm{Si\left(OH\right)_4}  &\mathrm{\rightarrow SiO_2 + 2H_2O}
\label{eq:oxidation}
\end{align}

where ($\mathrm{\nu_{VB}}$) represents the number of charge-carriers transferred
through valence-band processes and $\mathrm{(4-\nu_{VB})}$ the number
of charge-carriers transferred through conduction-band processes. The
second reaction is solely chemical, i.e. no charge-carriers are
transferred for the reaction. It has to be noted that the initial
charge transfer is always a valence-band process rendering
illumination necessary for the reaction to occur at n-type doped
silicon samples. The illumination also limits the total current, which
is a likely source of the non-linear global coupling \cite{Matsumura_JEC_1983}.

The generated oxide is etched away
by the fluoride species present in the electrolyte, e.g. HF \cite{Cattarin_JES_2000}, in
another solely chemical process:

\begin{equation}
  \mathrm{SiO_2 + 6HF \rightarrow SiF_6^{2-} + 2H^+ + 2H_2O}
\label{eq:etching}
\end{equation}

As silicon oxidation and the etching of silicon oxide have
opposite effects on the oxide-layer thickness, a steady state can be reached
for suitable experimental conditions. Already in the 1950s it was
found that the system can also be oscillatory, which
has drawn a lot of attention since then (for a review see
chapter 5 in ref.~\cite{Zhang_2001}).  
The current oscillations are accompanied by an oscillating oxide-layer
thickness with an amplitude in the nm-range \cite{Blackwood_ElectrochimActa_1992, Chazalviel_JES_1998,
  Yahyaoui_JES_2003, Chazalviel_ElectrochimActa_1992, Aggour_JEC_1995}. 

To investigate the spatial distribution of the oxide-layer
thickness during the oscillations we use spatially resolved
ellipsometric imaging, a technique first established by Rotermund
\textit{et. al.} \cite{Rotermund_Science_1995}, with a setup
described in the Appendix. The elliptical
polarization of a light beam is distorted
upon reflection from the working electrode surface by the silicon
oxide layer and these distortions are translated into a two-dimensional
representation of the oxide-layer thickness on the surface.

It was found that spatial pattern formation with a rich
variety of different patterns occurs on n-type doped silicon samples
at intermediate illumination intensities \cite{Miethe_PRL_2009,
  Miethe_PhDTh}.  An external resistor in series with the working
electrode acts as an additional linear global coupling
\cite{Krischer_ElectrochimActa_2003}.

\begin{figure}[h]
  \centering
  \includegraphics[width=85mm]{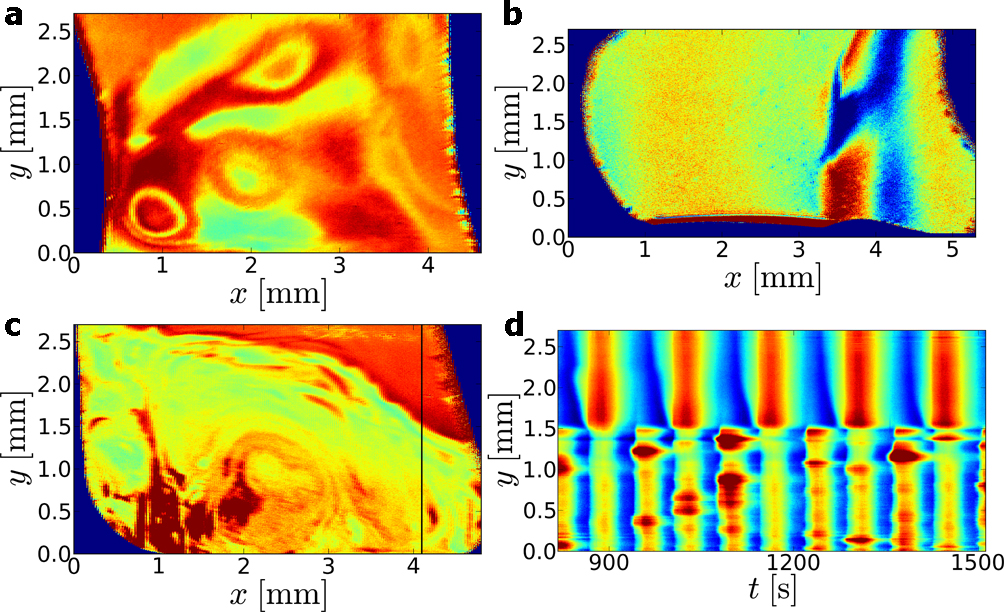}
  \caption{Spatio-temporal evolution of the oxide-layer thickness
    during the electrodissolution of silicon:
    two-phase clusters, sub-clustering and chimera state. Shown are
  snapshots, colours indicate the
  thickness of the oxide layer, $x$ and $y$ represent spatial
  coordinates and $t$ denotes time. (\textbf{a}) Two-phase cluster state, where
  both parts oscillate uniformly with a phase difference to the
  respective other. (\textbf{b}) The oxide-layer thickness exhibits
  sub-clustering: a stripe of two-phase clusters is embedded in an
  otherwise uniformly oscillating background. The clusters in the
  stripe oscillate at half the frequency of the background
  oscillation. (\textbf{c}) Chimera state: the
  coexistence of synchrony and asynchrony is apparent. (\textbf{d})
  Cut along $y$ (black line in (\textbf{c})) showing the sharp separation into 
  coherent and incoherent regions. For experimental parameters see
  the Appendix.}
\label{fig:Exp_combined}
\end{figure}

In Fig.~\ref{fig:Exp_combined}a-c we present experimentally measured
snapshots of the oxide-layer thickness. Consistent with the theory,
Fig.~\ref{fig:W_real}a, the case of two-phase clusters is shown in
Fig.~\ref{fig:Exp_combined}a.

As visible in Fig.~\ref{fig:Exp_combined}b, we also observe a
subclustering in the experiment as in Fig.~\ref{fig:W_real}b. A stripe exhibiting two-phase clusters is
embedded in an otherwise homogeneous region, which oscillates twice as
fast as the two-phase clusters. 

\begin{figure}[h]
  \centering
  \includegraphics[width=85mm]{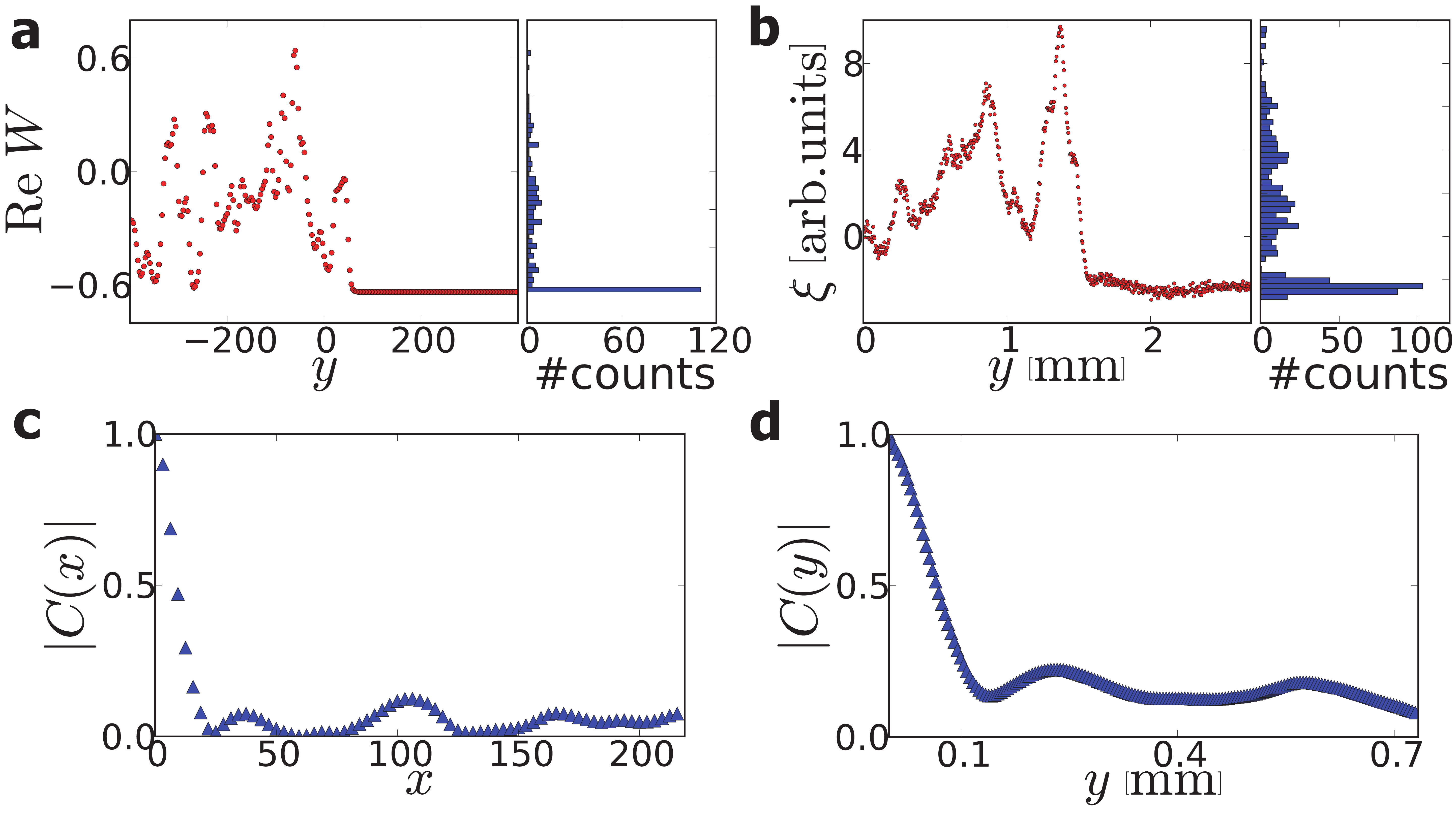}
  \caption{Comparison of theoretical and experimental chimera
    states. The one-dimensional spatial profiles in theory (\textbf{a}) and
    experiment (\textbf{b}) are in excellent agreement. Furthermore, both
    correlation functions $\left| C(x) \right|$ (for details see
    text) exhibit a fast drop to nearly zero. This shows the fast decrease of
    spatial correlations in theory (\textbf{c}) and experiment (\textbf{d}).}
  \label{fig:comp_profiles_C}
\end{figure}

Finally, and most remarkably, also the spontaneous formation of
a two-dimensional chimera state occurs in the experiments,
Figs.~\ref{fig:Exp_combined}c and d. As apparent in the snapshot
(Fig.~\ref{fig:Exp_combined}c)
and the time evolution of a one-dimensional cut (Fig.~\ref{fig:Exp_combined}d) the upper right corner of the electrode constitutes the
synchronized region, whereas the remaining part displays turbulent
dynamics. A one-dimensional snapshot of the oxide-layer thickness in Fig.~\ref{fig:comp_profiles_C}b,
with corresponding distribution visualized by a histogram, shows the
strong variations in the incoherent region.

We found the coexistence of synchrony and incoherence for several experimental
parameters. For sufficiently long measurement times, we observed a
transient nature. On the contrary, for all considered
simulation durations the chimera state remains stable in the ensemble
of Stuart-Landau oscillators, Eq.~\eqref{eq:SL_ensemble}, and in the MCGLE,
Eq.~\eqref{eq:MCGLE} for the given parameter values.

We have to point out that, as in the simulations, nothing is imposed
onto the system to introduce the splitting into two domains. This
separation arises solely from the intrinsic dynamics. We remark as
well that there is no Turing-Hopf bifurcation \cite{DeWit_PRE_1996} or an analogous
situation that would trigger the splitting. Furthermore,
great care was taken to assure that the experimental conditions are
spatially uniform. To this end the electrolyte is stirred continuously
and the counter electrode is placed symmetrically in front of the
silicon working electrode.

Finally we make a direct comparison of the theoretical and
experimental spatial profiles of the real part of $W$ in Fig.~\ref{fig:comp_profiles_C}a and the
oxide-layer thickness $\xi$ in Fig.~\ref{fig:comp_profiles_C}b,
respectively. These plots show an excellent qualitative agreement.

Furthermore, we quantified the incoherence in the turbulent regions of
the chimera state: we calculated the correlation function $C(x,t) =
\left< \tilde W(x,t) \tilde W^*(0,0) \right>_{x',t'} / \left< \left|
    \tilde W(0,0) \right|^2 \right>_{x',t'}$ (the asterisk denotes complex
conjugation and the average is performed over space and time) in a cut
in the incoherent region for both theory and experiment, where $\tilde
W$ is obtained by subtracting the average of this cut. From the experimental data
$W(x,t)$ was obtained via a Hilbert transformation. The resulting $\left| C(x)
\right| \equiv \left| C(x,0) \right|$ is shown in
Figs~\ref{fig:comp_profiles_C}c (theory) and d (experiment). As seen in
the figures, $\left| C(x) \right|$ drops very fast to approximately zero,
demonstrating that after this distance the individual oscillators
behave uncorrelated. Note that the fluctuations of $\left| C(x) \right|$ are due to
the finiteness of the sample.
We point out that neither in the theoretical nor in the experimental
profiles amplitude defects are present. This situation contrasts with
the so-called localized turbulence found under linear global coupling
\cite{Battogtokh_PhysicaD_1997}.


\section{Conclusions}

In this article we demonstrate that two-dimensional chimera states and other spatial
symmetry breakings may spontaneously occur in systems with nonlinear global
coupling, both theoretically and experimentally. Simulations of an
ensemble of Stuart-Landau oscillators, coupled solely via the
nonlinear global coupling, provide evidence that a nonlocality of the
coupling is dispensable for the formation of chimera states. The
spontaneity of the formation of chimeras is astonishing and affirms
earlier theoretical observations \cite{Omelchenko_PRL_2008}.

The theoretical description is very general and a nonlinear global
coupling seems to be essential for the modelling of subharmonic cluster
patterns, where the clusters oscillate at a lower frequency than
the homogeneous mode \cite{GarciaMorales_PRE_2010}.
Subharmonic clusters were observed in a number of experiments
\cite{Miethe_PRL_2009, Varela_PCCP_2005, Vanag_JPCA_2000, Kim_Science_2001, Pollmann_CPL_2001},
suggesting that also the type of
symmetry breaking described here, especially the chimera state, may occur spontaneously in many
physical and chemical systems. Furthermore, as shown in
\cite{Kiss_PRL_2005}, the proposed nonlinearity of the global coupling
may also arise effectively in systems of linearly coupled relaxational oscillators.



%
%

%

\begin{acknowledgments}
We thank Andreas Heinrich and Martin Wiegand for assistance on the
experiments and Moritz M\"{u}ller for his work on the simulation program.
Financial support from the \textit{Deutsche Forschungsgemeinschaft}
(grant no. KR1189/12-1),
the \textit{Institute for Advanced Study - Technische Universit\"{a}t
  M\"{u}nchen} funded by the German Excellence Initiative and the cluster of excellence \textit{Nanosystems
  Initiative Munich (NIM)} is gratefully acknowledged.
\end{acknowledgments}

\appendix
\section{}
\subsection{Simulations of the ensemble of Stuart-Landau oscillators}
We numerically solve Eq.~(1) in the main text using an implicit
Adams method with timestep $dt = 0.01$. The system consists of $N =
1000$ oscillators, initialized with random real numbers (with the
condition on their average fulfilled). Note that the equation is dimensionless.

\subsection{Simulations of the modified complex Ginzburg-Landau equation}
Simulations of Eq.~(2) in the main text are carried out using a
pseudospectral method and an exponential time stepping algorithm
\cite{CoxMatthews_JCompPhys_2002}. We use 512x512 Fourier modes and a
system size of $L = 800$. Note that the equation is
dimensionless. The system is initialized with a two-dimensional
circular perturbation and additional noise. The dynamics is analyzed between $t = 500$ and $t
= 1000$ and we use a computational timestep of $\Delta t = 0.05$.

\subsection{Experiments}
\begin{figure}[t]
  \centering
  \includegraphics[width=87mm]{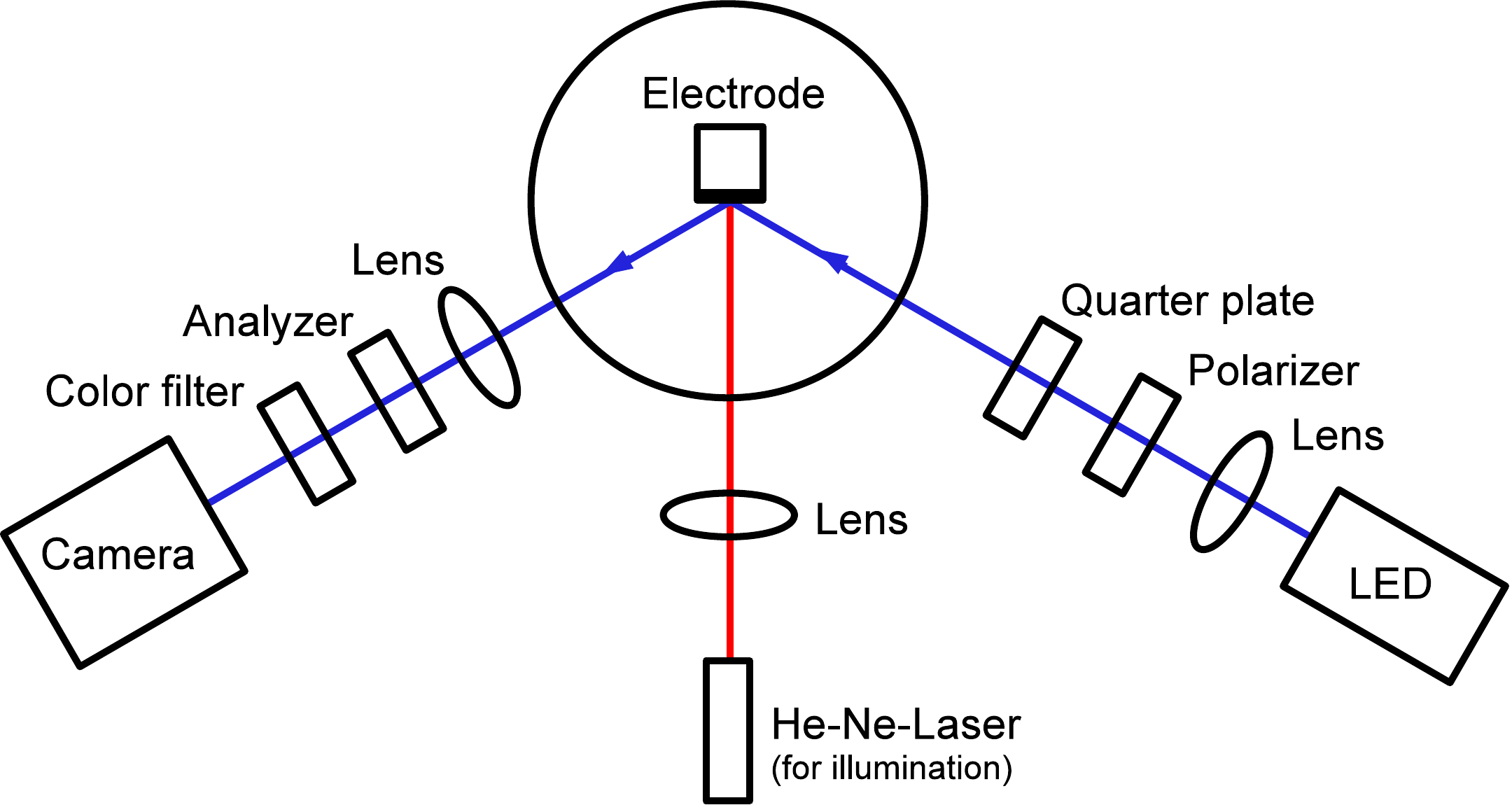}
  \caption{Optical setup of the custom-made ellipsometric
      microscope. The blue (dark gray) line represents the light path for the
    imaging and the red (light gray) line for the illumination.}
\label{fig:Exp_setup}
\end{figure}

The experiments are carried out in a custom made PTFE three electrode electrochemical cell
with a monocrystalline n-Si ((111) surface, 3-5 $\Omega$cm) working electrode, a Hg$|$Hg$_2$SO$_4$
reference electrode and a ring-shaped platinum counter electrode
placed symmetrically in front of the working electrode
\cite{Miethe_PRL_2009}. The working electrode has an ohmic aluminum
back contact annealed at 250$\mathrm{^\circ C}$ for 15 min
and otherwise prepared
as described in an earlier work \cite{Schoenleber_ChemPhysChem_2012}. We use a
NH$_4$F solution as electrolyte, adjust the pH value by adding
H$_2$SO$_4$ and stir with a magnetic
stirrer at about 10 Hz. For illumination a He-Ne Laser (633 nm) is
used, whose intensity $I$ is tuned by a polarizer. 
For all experiments a voltage of 8.65 V vs. SHE (Standard Hydrogen Electrode) is applied and the current
response is recorded. For the spatially resolved ellipsometric imaging, elliptically polarized light (LED, 470 nm) 
is reflected from the working-electrode surface at an angle of
$70^\circ$, close to the Brewster angle, which is to maximize the contrast. The reflected light then passes another polarizer, 
that converts changes of the polarization upon interaction with the
surface into intensity changes, and is imaged on a CCD chip
(640$\times$480 pixels). For a schematic setup see Fig.~\ref{fig:Exp_setup}.

The data are then recorded
using a suitable \texttt{LabVIEW} program and analyzed with
\texttt{MATLAB}. The parameters varied in the experiments are: the
concentration of $\mathrm{NH_4F}$, $\mathrm{\left[ NH_4F \right]}$, the
surface area of the working electrode, $A$, the external resistance,
$R_{\mathrm{ext}}$, and the illumination intensity, $I$. Values read:
$\mathrm{\left[ NH_4F \right]} = 35 \ \mathrm{mM}$, pH $=1$, $A = 22.73 \ 
\mathrm{mm^2}$, $R_{\mathrm{ext}} = 40 \ \mathrm{k\Omega}$, $I \simeq
0.7 \ \mathrm{mW/cm^2}$ (two-phase clusters), $\mathrm{\left[ NH_4F
  \right]} = 50 \ \mathrm{mM}$, pH $=2.3$, $A = 23.06 \ 
\mathrm{mm^2}$, $R_{\mathrm{ext}} = 0 \ \mathrm{\Omega}$, $I \simeq 1
\ \mathrm{mW/cm^2}$ (sub-clustering) and $\mathrm{\left[ NH_4F
  \right]} = 50 \ \mathrm{mM}$, pH $=3$, $A = 22.42 \ 
\mathrm{mm^2}$, $R_{\mathrm{ext}} = 0 \ \mathrm{\Omega}$, $I \simeq
0.5 \ \mathrm{mW/cm^2}$ (chimera).

%

\end{document}